\documentclass[aps,prx,twocolumn,amsmath,amssymb,floatfix]{revtex4-1}
\usepackage{tabularx}
\usepackage{bm}
\usepackage{euscript}
\usepackage{epsfig,psfrag,subfigure}
\usepackage{graphicx}
\usepackage{color}
\usepackage{amsfonts}
\usepackage{exscale}
\usepackage{amsbsy}
\usepackage{multirow}

\def\avg#1{\left\langle#1\right\rangle}

\def\abs#1{\left|#1\right|}

\def\trace#1{{\rm Tr}\left[#1\right]}

\def\be{\begin{equation}}       \def\ee{\end{equation}}
\def\bea{\begin{eqnarray}}      \def\eea{\end{eqnarray}}
\def\ba{\begin{array} }
\def\ea{\end{array} }

\def\nn{\nonumber}

\def\=>{\Rightarrow}
\def\>{\rightarrow}
\def\A{\uparrow}
\def\V{\downarrow}
\def\Eq#1{Eq.~(\ref{#1})}
\def\Fig#1{Fig.~\ref{#1}}

\renewcommand{\v}[1]{{\bf #1}}

\newcommand{\s}{{\sigma}}

\renewcommand{\>}{\rangle}

\begin{document}

\title{Topological superconductivity in the doped chiral spin liquid on the triangular lattice}
\author{Yi-Fan Jiang}
\affiliation{Stanford Institute for Materials and Energy Sciences, SLAC National Accelerator Laboratory and Stanford University, Menlo Park, CA 94025, USA}

\author{Hong-Chen Jiang}
\email{hcjiang@stanford.edu}
\affiliation{Stanford Institute for Materials and Energy Sciences, SLAC National Accelerator Laboratory and Stanford University, Menlo Park, CA 94025, USA}

\begin{abstract}
It has long been proposed that doping a chiral spin liquid (CSL) or fractional quantum Hall state can give rise to topological superconductivity. Despite of intensive effort, definitive evidences still remain lacking. We address this problem by studying the $t$-$J$ model supplemented by time-reversal symmetry breaking chiral interaction $J_\chi$ on the triangular lattice using density-matrix renormalization group with a finite concentration $\delta$ of doped holes. It has been established that the undoped, i.e., $\delta$=0, system has a CSL ground state in the parameter region $0.32\le J_\chi/J \le 0.56$. Upon light doping, we find that the ground state of the system is consistent with a Luther-Emery liquid with power-law superconducting and charge-density-wave correlations but short-range spin-spin correlations. In particular, the superconducting correlations, whose pairing symmetry is consistent with $d\pm id$-wave, are dominant at all hole doping concentrations. Our results provide direct evidences that doping the CSL on the triangular lattice can naturally give rise to topological superconductivity.
\end{abstract}
\date{\today}

\maketitle

Quantum spin liquids (QSLs) are exotic phases of matter that exhibit various novel features associated with their topological character and support fractional excitations.\cite{balents2010,Lee2006,Zhou2016a,Savary2016,Broholm2019} QSLs have attracted broad interest for several decades as an insulating phase with preexisting electron pairs, such that it might naturally yield high temperature superconductivity upon light doping with holes.\cite{Broholm2019,Anderson1987,Rokhsar1988,Wen1996,Lee2007,Fradkin2015,Laughlin1988,Laughlin1988a,Wen1989,Kalmeyer1987,Venderley2019,Venderley2019a} More broadly, it has been proposed that a host of behaviors of highly correlated electronic systems can be best understood from the perspective of doped QSLs.\cite{Senthil2003,Patel2016} It has also been believed that simple models such as Hubbard model and its strong-coupling limit, the $t$-$J$ model, may contain enough physics to explain the appearance of superconductivity and other phases by doping a QSL.\cite{Lee2006} However, despite intense efforts during past decades, definitive evidences showing that doping a QSL leads to superconductivity remain still deficient. This is partially due to the fact that the realization of QSLs is a great challenge to physicists and candidate materials are rare.

One of the most promising spin liquid candidates is the spin-1/2 antiferromagnet with both the nearest-neighbor (NN) $J$ and next-nearest-neighbor $J_2$ Heisenberg interactions on the triangular lattice. A number of simulations provide strong evidences that its ground state is a QSL in the parameter region $0.07\lesssim J_2/J\lesssim 0.15$.\cite{Zhu2015,Hu2015,Zheng2015,Saadatmand2016,Gong2019,Hu2019,Gong2017} In addition, a new spin liquid phase has been found in the spin-1/2 triangular lattice Heisenberg model with additional time-reversal symmetry (TRS) breaking scalar chiral interaction $J_\chi$, i.e., $J$-$J_\chi$ terms in Eq.(\ref{eq:model}). Recent studies provide strong evidences that a novel chiral spin liquid (CSL) phase can be stabilized by the scalar chiral $J_\chi$ interaction in the prameter region $0.32\le J_\chi/J \le 0.56$.\cite{Gong2017,Hu2016,Wietek2017} Moreover, the CSL may also be realized in the Hubbard model at half-filling on the triangular lattice in the parameter region $8\le U/t\le 11$.\cite{Szasz2018} The universal properties of this CSL state are captured by boson $\nu=1/2$ Laughlin state as proposed by Kalmeyer and Laughlin.\cite{Kalmeyer1987} 
%Besides the triangular lattice, the CSL can also be realized on the kagome lattice models.\cite{He2014,Gong2014,Bauer2014,Wietek2015}
Physically, this scalar chiral term $J_\chi\sim \Phi t^3/U^2$ can be obtained in the Hubbard model at half-filling from the $t/U$ expansion to the second order with large $U$ and an external magnetic field\cite{Sen1995,Motrunich2006}. Here $U$ denotes the Hubbard repulsion and $\Phi$ is the magnetic flux through each triangle. The chiral interaction can also be realized in cold atom system.\cite{Aidelsburger2013,Miyake2013}

\begin{figure}[tb]
    \centering
    \includegraphics[width=\linewidth]{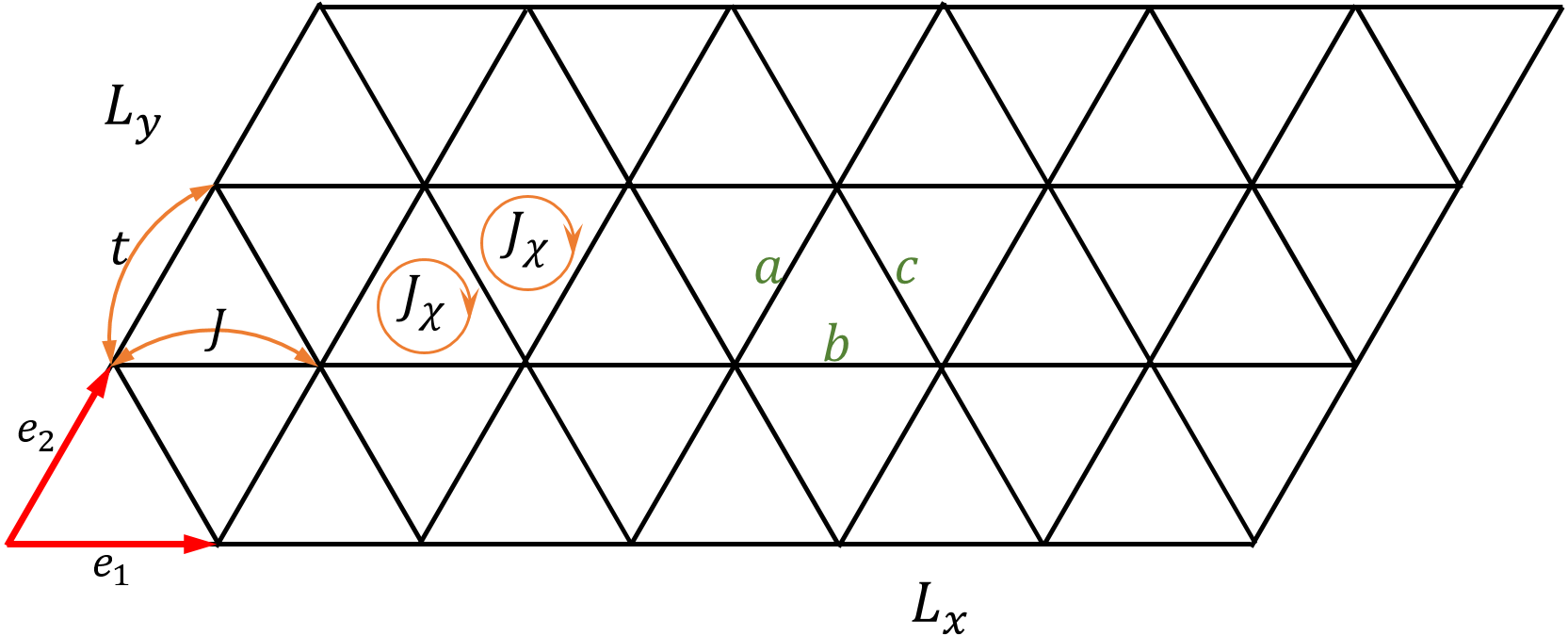}
    \caption{The $t$-$J$-$J_\chi$ model in Eq.(\ref{eq:model}) on the triangular lattice. The scalar chiral interaction $J_\chi$ in each triangle has the same chirality following the clockwise order. Periodic and open boundary conditions are imposed respectively along the directions specified by the lattice basis vectors, $\v{e}_2$ and $\v{e}_1$. $L_x$ and $L_y$ are the number of sites in the $\v{e}_1$ and $\v{e}_2$ directions, respectively. $a$, $b$ and $c$ label the three different bonds.}
    \label{fig:lattice}
\end{figure}

As the QSL has been realized on the spin-1/2 antiferromagnet on the triangular lattice, a natural question is that will dope this QSL yield superconductivity? Quasi-one-dimensional systems such as cylinders (depicted in Fig.\ref{fig:lattice}) have become an important starting point to answer the question. The cylinder can be viewed as one-dimensional (1D) but has essential degrees of freedom that allow for two-dimensional (2D) characteristics to emerge. According to the Mermin-Wagner theorem, a superconducting state that can be realized in quasi-1D systems such as cylinders has quasi-long-range superconducting (SC) correlation. The Luther-Emery (LE) liquid is an example of this,\cite{Luther1974,Dolfi2015,Jiang2018,Jiang2018a,Jiang2019} which has one gapless charge mode with quasi-long-range SC and charge-density-wave (CDW) correlations, but the spin-spin correlation is short-ranged. Recent DMRG calculations have found strong evidences that the LE liquid state can be realized in the lightly doped TRS preserving QSL of the $J$-$J_2$ model on the triangular lattice.\cite{Jiang2019a}

Topological superconductivity is another novel state of matter\cite{Senthil1999,Read2000,Kitaev2001,Sato2017}, which has attracted tremendous interest recently and has potential application to topological quantum computation.\cite{Kitaev2003,Nayak2008} It has been long proposed that doping a CSL or fractional quantum Hall state can give rise to topological superconductivity.\cite{Laughlin1988, Laughlin1988a,Kalmeyer1987, Wen1989} However, whether this is the case still needs to be confirmed. In this paper, we directly address this question by studying the $t$-$J$ model supplemented by chiral $J_\chi$ interaction defined in Eq.(\ref{eq:model}). Through large-scale density-matrix renormalization group (DMRG) simulations, we provide direct evidences that doping the CSL state on the triangular lattice will naturally yield topological superconductivity with $d\pm id$-wave pairing symmetry. In particular, the SC correlations are dominant in all cases studied. To the best of our knowledge, this is the first numerical demonstration of topological superconductivity from doping a CSL.

%==Fig.1==

{\bf Model and Method: } We employ the DMRG with spin SU(2) symmetry \cite{White1992,McCulloch2002,Schollwock2011} to investigate the ground state properties of the hole-doped $t$-$J$ model with additional TR symmetry breaking chiral interaction on the triangular lattice. The model Hamiltonian is defined as%
\bea
H=&-&t \sum_{\avg{ij}}( c_{i\s}^\dagger c_{j\s} + h.c.)+J\sum_{\avg{ij}} (\v{S}_i \cdot \v{S}_j -\frac14 \hat{n}_i \hat{n}_j)\nn\\ 
&+&J_\chi \sum_{\bigtriangleup/\bigtriangledown} (\v{S}_i\times \v{S}_j)\cdot \v{S}_k ,
\label{eq:model}
\eea
where $c_{i\s}^\dagger$ denotes the electron creation operator with spin $\sigma$ on site $i$. $\v{S}_i$ is the spin operator and $\hat{n}_i=\sum_\s c_{i\s}^\dagger c_{i\s}$ is the electron number operator. $t$ and $J$ denote the NN electron hopping integral and spin interaction, respectively. The local Hilbert space is constrained by the no-double occupancy condition, $n_i \le 1$. The chiral interaction $J_\chi$ has the same amplitude for all the up and down triangles, and the three sites $i$, $j$ and $k$ in each triangle follow the clockwise order shown in \Fig{fig:lattice}. At half filling, \Eq{eq:model} reduces to the $J$-$J_\chi$ model, which has a stable gapped CSL phase in the parameter region $0.32\le J_\chi/J \le 0.56$.\cite{Gong2017,Hu2016,Wietek2017}

The lattice geometry used in our simulation is depicted in \Fig{fig:lattice}, where $\v{e}_1$ and $\v{e}_2$ denote the two basis vectors of the triangular lattice. We consider triangular cylinder with periodic (open) boundary condition in the $\v{e}_2$ ($\v{e}_1$) direction. We focus on cylinders with width $L_y$ and length $L_x$, where $L_y$ and $L_x$ are the number of sites along the $\v{e}_2$ and $\v{e}_1$ directions, respectively. The hole doping concentration away from half-filling is defined as $\delta=N_h/N$, where $N=L_x\times L_y$ is the total number of lattice sites and $N_h$ is the number of doped holes. For the present study, we focus on the lightly-doped case with $\delta\le 8.33\%$. We set $J$=1 as an energy unit and consider $t$=3 and $J_\chi$=0.4, which is deep in the CSL phase of the $J$-$J_\chi$ spin model. Our results also hold for other $J_\chi$ which are provided in the Supplemental Material (SM). In this paper, we focus on cylinders of width $L_y$=4 with length up to $L_x$=96. We keep up to $m=6000$ SU(2) states (effectively $18000$ U(1) states) to obtain accurate results with truncation error $\epsilon\leq 10^{-6}$. This leads to excellent convergence for our results when extrapolated to $m=\infty$ limit.

%==Fig.2==
\begin{figure}[tb]
    \centering
    \includegraphics[width=\linewidth]{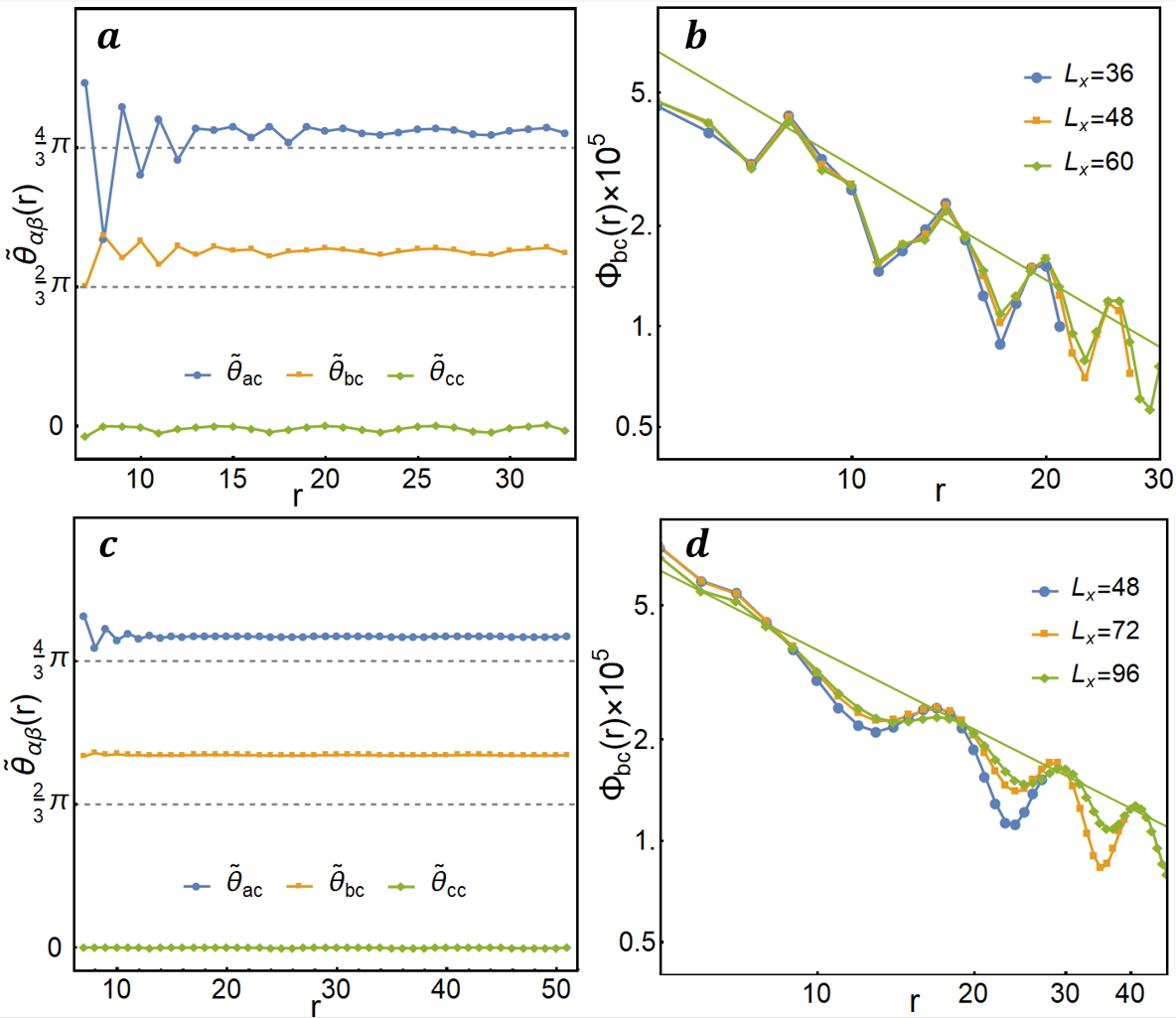}
    \caption{Superconducting correlations. The pattern of phase $\tilde \theta_{\alpha\beta}$ for $L_y=4$ cylinders at (a) $\delta=8.33\%$ and (c) $\delta=4.17\%$. The amplitude of the superconducting correlations $\abs{\Phi(r)}$ on $L_y=4$ cylinders at doping levels (b) $\delta=8.33\%$ and (d) $\delta=4.17\%$ on double-logarithmic scales. Solid lines denote power-law fitting $|\Phi(r)|\sim r^{-K_{sc}}$.} 
    \label{fig:paircor}
\end{figure}

%==Principal results==
\textbf{Principal results:} %
We find that the system exhibits power-law SC correlations for $L_y$=4 cylinders in the lightly doped region $\delta\leq 8.33\%$. As $L_x\gg L_y$, the system can be thought of as quasi-1D, the long-range superconductivity expected in two dimensions will decay as a power-law due to the Mermin-Wagner theorem. The SC pair-pair correlation function $\Phi_{\alpha\beta}(r)$, defined in \Eq{eq:phi}, can be characterized by a power-law (see \Fig{fig:paircor}b and d), with the appropriate Luttinger exponent $K_{sc}$. % shown in \Fig{fig:exponent}. 
Different with doping a TRS preserving QSL,\cite{Jiang2019a} the pairing symmetry of the SC correlations, visualized by the phase $\theta_{\alpha\beta}$ (see \Fig{fig:paircor}a and c), is consistent with the $d\pm id$-wave. Therefore, our results show that doping the CSL on the triangular lattice will yield topological $d\pm id$-wave superconductivity.

\begin{figure}[tb]
    \centering
    \includegraphics[width=\linewidth]{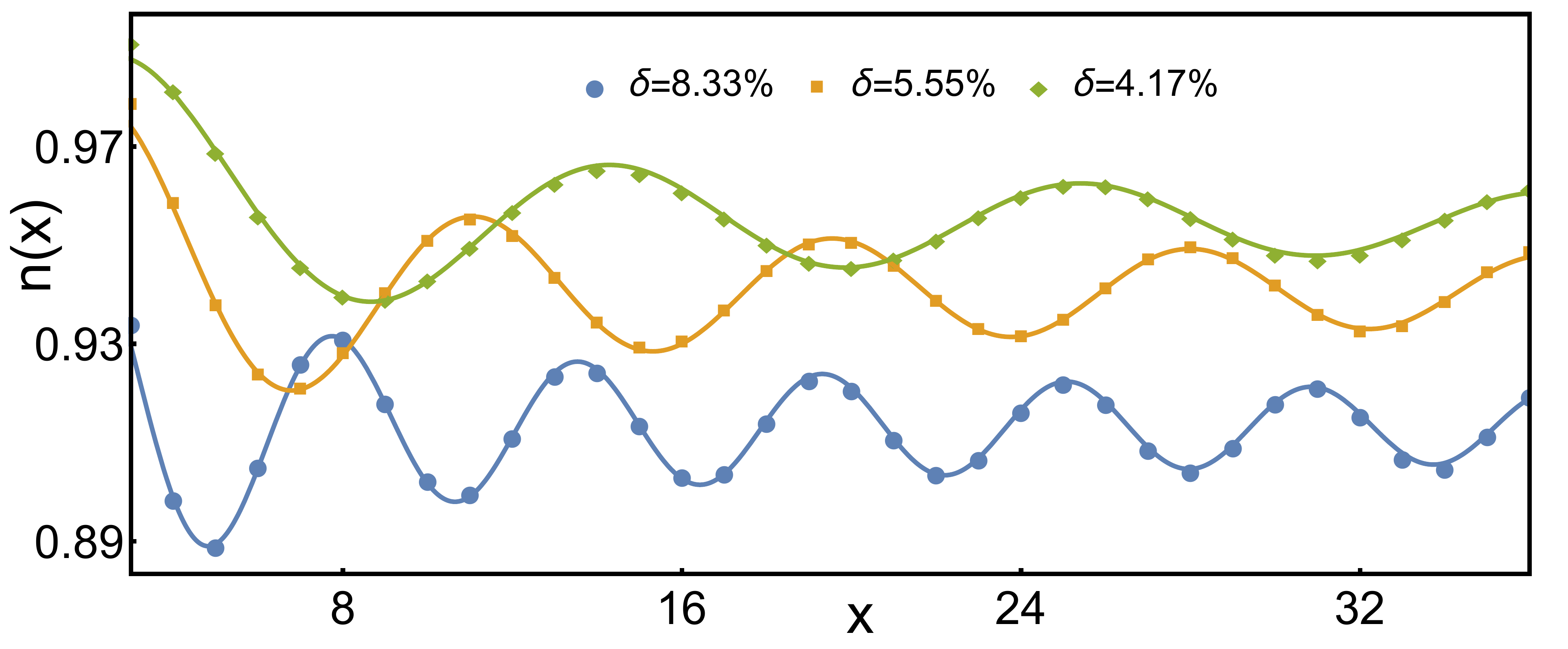}
    \caption{Charge density profiles $n(x)$ for $L_y=4$ cylinders of length $L_x=72$ at doping levels $\delta=4.17\%\sim8.33\%$. Solid lines are the fits from Friedel oscillation using Eq.(\ref{eq:friedel}).}
    \label{fig:density}
\end{figure}

Similar with the SC correlation, the CDW correlation also decays with a power-law corresponding to a local pattern of ``partially filled'' charge stripes. The wavelength of the CDW, i.e., the spacings between two adjacent charge stripes, is $\lambda=1/2\delta$. This charge stripe has an ordering wavevector $Q=4\pi\delta$ with half a doped hole per each CDW unit cell (see Fig.\ref{fig:density}). This is similar to the lightly doped $t$-$J$ and Hubbard models on $L_y=4$ square cylinder with second-neighbor interactions,\cite{Jiang2018,Jiang2018a,Jiang2019} and lightly doped TR-invariant QSL on the triangular lattice.\cite{Jiang2019a} 

\begin{figure}[tb]
    \centering
    \includegraphics[width=\linewidth]{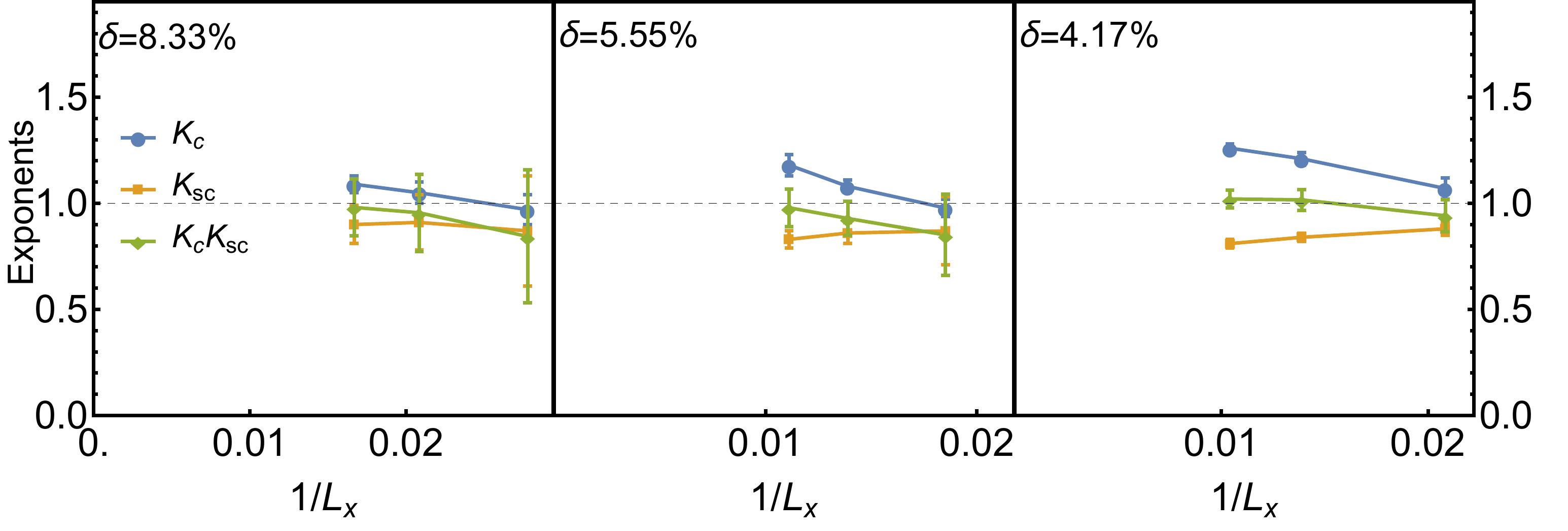}
    \caption{The extracted Luttinger exponents $K_{sc}$, $K_c$ and their product $K_{sc}K_{c}$ on $L_y=4$ cylinders as a function of $L_x$ at different doping levels $\delta$. Dashed lines are guides for eyes.} 
    \label{fig:exponent}
\end{figure}

\begin{figure}[tb]
    \centering
    \includegraphics[width=\linewidth]{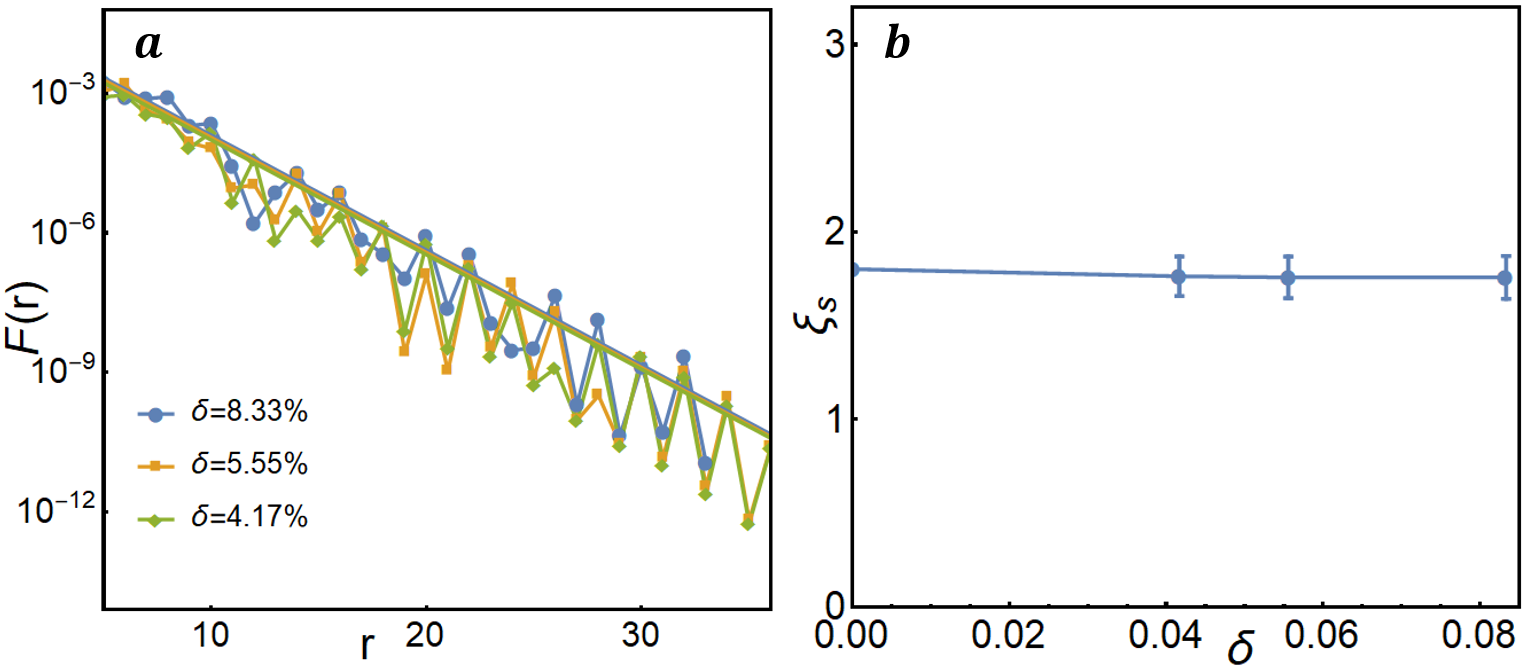}
    \caption{Spin-spin correlations. (a) Examples of spin-spin correlations $F(r)$ on $L_y=4$ cylinders of length $L_x=72$ on the semi-logarithmic scale at different $\delta$. Solid lines denote exponential fitting $F(r)\sim e^{-r/\xi_s}$.  (b) Spin-spin correlation length $\xi_s$ as a function of $\delta$. Error bars denote the numerical uncertainty.} 
    \label{fig:spincor}
\end{figure}

\begin{figure}[tb]
    \centering
    \includegraphics[width=\linewidth]{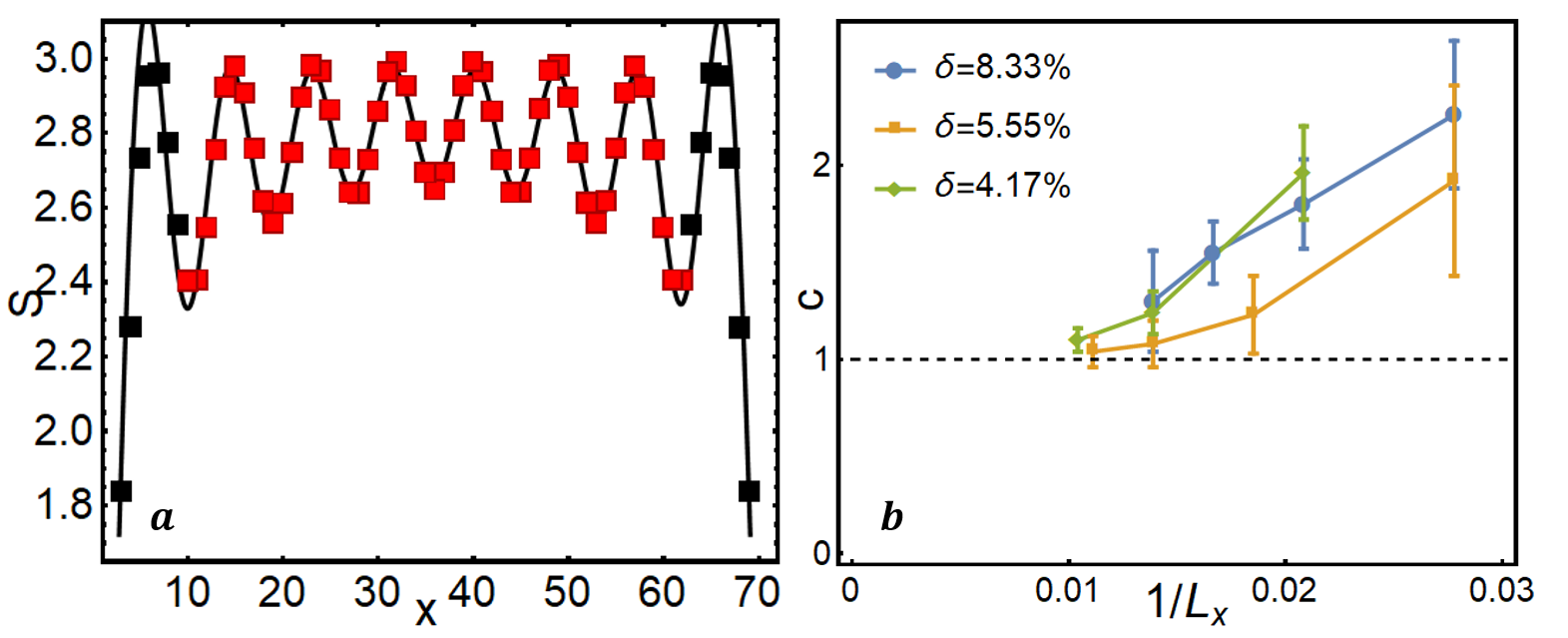}
    \caption{Entanglement entropy and central charge. (a) Entanglement entropy $S(x)$ on $L_y=4$ cylinder of length $L_x=72$ at doping $\delta=5.55\%$. A few data points in black are removed to minimize the boundary effect. (b) Central charge $c$ at different $\delta$ as a function of $L_x$. Error bars denote the numerical uncertainty.}
    \label{fig:entropy}
\end{figure}

The Luttinger exponents $K_{sc}$ and $K_c$ extracted from the SC and CDW correlations using Eq.(\ref{eq:ksc}) and Eq.(\ref{eq:friedel}), respectively, are shown in Fig.\ref{fig:exponent}. It is worth emphasizing that $K_{sc}<K_c$ for all the cases in the present study, which directly demonstrates the dominance of the SC correlations. Contrary to SC and CDW correlations, both single-particle and spin-spin correlations (see Fig.\ref{fig:spincor} and SM) decay exponentially at long distances. This suggests that the ground state of the lightly doped CSL is consistent with the LE liquid in the quasi-1D limit, which is characterized by one gapless charge model with quasi-long-range SC and CDW correlations but short-range spin-spin correlation. Moreover,our results show that the relation $K_c K_{sc}=1$, which is expected from the LE liquid\cite{Luther1974}, also holds within the numerical uncertainty and finite-size effect (see Fig.\ref{fig:exponent}). Further support from the central charge $c$, where a clear saturation to $c=1$, i.e, one gapless charge mode, has been observed in the $L_x\gg L_y$ limit (see Fig.\ref{fig:entropy}).

%==Fig.3==

%==Topological superconductivity==
{\bf Topological superconductivity: }%
It has long been proposed that doping a CSL, or equivalently the $\nu=1/2$ bosonic fractional quantum Hall state\cite{Kalmeyer1987,Laughlin1988, Laughlin1988a}, will give rise to topological superconductivity. To address this, we have calculated the equal-time SC pair-pair correlation function. As the ground state of the system with even number of doped holes is always found to have spin 0, we focus on spin-singlet SC correlation, which is defined as%
\bea
\Phi_{\alpha\beta}(r)=\frac{1}{L_y}\sum_{y=1}^{L_y}\avg{\Delta_\alpha^\dagger(x_0,y)\Delta_\beta(x_0+r,y)}. \label{eq:phi}
\eea
Here $\Delta_\alpha^\dagger(i)=\frac{1}{\sqrt{2}}(c_{i\A}^\dagger c_{i+\alpha\V}^\dagger- c_{i\V}^\dagger c_{i+\alpha\A}^\dagger)$ is the spin-singlet pair-field creation operator, and $\alpha=a$, $b$ and $c$ denotes the bond orientation (see Fig.\ref{fig:lattice}). ($x_0,y$) is the reference site which is located at $x_0=L_x/4$, and $r$ is the distance between two bonds in the $\v{e}_1$ direction.

As the ground state of the system is always a spin-singlet, the pairing symmetry of the SC correlation will be consistent with either $s$-wave or $d$-wave, including nematic $d$-wave and $d\pm id$-wave.\cite{Raghu2010} If the pairing symmetry is $s$-wave, or nematic $d$-wave which breaks the lattice rotational symmetry, then the SC correlation $\Phi_{\alpha\beta}(r)$ should be purely real. However, this is inconsistent with our results. This makes the $d\pm id$-wave pairing symmetry the only choice, which is indeed consistent with our results. To show this, it is convenient to rewrite the SC correlation as $\Phi_{\alpha\beta}(r)=\abs{\Phi_{\alpha\beta}(r)}e^{i \theta_{\alpha\beta}(r)}$, where $|\Phi_{\alpha\beta}(r)|$ and $\theta_{\alpha\beta}(r)$ denote its amplitude and phase,  characterizing its decaying behavior and pairing symmetry, respectively. Numerically, the pairing symmetry can be determined by the pattern of phases $\theta_{\alpha\beta}(r)$. For instance, if we take $a$-bond as a reference, the value of $\{\theta_{aa}(r), \theta_{ab}(r), \theta_{ac}(r)\}$ should converge to $\{0,4\pi/3,2\pi/3\}$ for $d$+$id$-wave or $\{0,2\pi/3,4\pi/3\}$ for $d$-$id$-wave. Examples are shown in Fig.\ref{fig:paircor}a and c.\footnote{Note that a modified $\tilde{\theta}$ (see SM) is used for $L_y=4$ cylinder}. Although a visible deviation and small spatial oscillation have been numerically observed, these could be attributed to the finite-size effect that the width $L_y=4$ cylinders are still relatively narrow. This is supported by our results on wider $L_y=6$ cylinders (see SM details), where both the deviations and spatial oscillations become significantly smaller. Therefore, our results are consistent with the expected $d\pm id$-wave pairing symmetry even the lattice geometry explicitly breaks the $C_3$ rotational symmetry.

At long distances, we find that the SC correlation is characterized by a power-law with the appropriate Luttinger exponent $K_{sc}$ defined by%
\bea
\abs{\Phi_{\alpha\beta}(r)} \propto r^{-K_{sc}}. \label{eq:ksc}
\eea
The extracted $K_c$ at three different doping levels are shown in Fig.\ref{fig:exponent}. The fact that the observed $K_{sc}< 1$ which holds for all the cases and decreases with $L_x$ clearly suggests that a smaller $K_{sc}$ with stronger SC correlations is expected in the long cylinder limit. This provide strong evidences for robust SC correlations with divergent SC susceptibility. Moreover, our results show that $K_{sc}<K_c$ for all the cases in the present study, which undoubtedly demonstrates the dominance of the SC correlations. More results are provided in the SM.

%==Fig.4==

%==Fig.5==

%==Charge density wave order==
{\bf Charge density wave order: }%
To describe the charge density properties of the system, we have also calculated the charge density profile $n(x)=\sum_{y=1}^{L_y}\langle \hat{n}(x,y)\rangle /L_y$, where $x$ is the rung index of the cylinder. Our results on the $L_y=4$ cylinders (see Fig.\ref{fig:density}) are consistent with the ``half-filled'' charge stripe, which is defined by the periodicity along the cylinder. The wavelength is doping dependent $\lambda=1/2\delta$ with half a doped hole in each CDW unit cell.

Similar with the SC correlations, the spatial decay of the CDW correlations is also dominated by a power-law with the Luttinger exponent $K_c$, which can be extracted by fitting the Friedel oscillation (charge density oscillation) induced by the open boundaries of the cylinders\cite{White2002}%
\bea
n(x)=n_0 + \delta n \cos(2k_F x + \phi)x^{-K_c/2}, \label{eq:friedel}
\eea
$\delta n$ denotes the non-universal amplitude, $\phi$ the phase shift, $n_0$ the background density and $k_F$ is the Fermi vector. The $K_c$ is shown in Fig.\ref{fig:exponent} for several doping concentrations as a function of $L_x$. For all cases, we find that $K_c>K_{sc}$ showing that the SC correlation is dominant while the CDW correlation is weaker. Similarly, $K_c$ can also be obtained by fitting the charge density-density correlation, which gives consistent results. Details are provided in the SM.

%==Spin-spin correlation==
{\bf Spin-spin correlation: }%
To describe the magnetic properties of the ground state of the system, we have calculated the spin-spin correlations $F(r)=\frac{1}{L_y}\sum_{y=1}^{L_y} |\avg{\v{S}_{x_0,y}\cdot \v{S}_{x_0+r,y}}|$. Here $\v{S}_{x,y}$ is the spin operator on site $i=(x,y)$, $(x_0,y)$ is the reference site with $x_0\sim L_x/4$ and $r$ is the distance between two sites along the cylinder. At half-filling, i.e., $\delta=0$, the CSL state is fully gapped with short-range spin-spin correlation $F(r)\sim e^{-r/\xi_s}$ where $\xi_s$ is the spin-spin correlation length. Upon light doping, we find that $F(r)$ still decays exponentially as shown in Fig.\ref{fig:spincor}, with a nearly doping independent correlation length $\xi_s\sim 1.5$ lattice spacings.
Therefore, we conclude that the spin-spin correlations are short-ranged with a finite gap in the spin sector, which is similar with the CSL at half-filling.

%==Fig.6==

%==Central charge==
{\bf Central charge: }%
Our results suggest that the ground state of the lightly-doped CSL is consistent the LE liquid, which has one gapless charge mode with central charge $c=1$. To show this, we calculate the von Neumann entropy $S(x)=-\trace{\rho_x \ln \rho_x}$, where $\rho_x$ is the reduced density matrix of the subsystem with length $x$. For critical systems in 1+1 dimensions described by the conformal field theory, it has been established\cite{Calabrese2004,Fagotti2011} that for an open system of length $L_x$%
\bea
S(x)&=&\frac{c}{6} \ln \big[\frac{4(L_x+1)}{\pi} \sin \frac{\pi(2x+1)}{2(L_x+1)}|\sin k_F|\big] \nn\\
&+& \frac{A\sin[k_F(2x+1)]}{\frac{4(L_x+1)}{\pi} \sin \frac{\pi(2x+1)}{2(L_x+1)}|\sin k_F|}+ B. \label{Eq-EE}
\eea
Here A and B are fitting parameters and $k_F$ is the Fermi momentum. Fig.\ref{fig:entropy}a shows an example of $S(x)$ for the $L_x=72$ cylinder, where a few data points in black are excluded in the fitting to minimize the boundary effect. The extracted $c$ is given in Fig.\ref{fig:entropy}b as a function $L_x$ at different $\delta$. It is clear that the central charge quickly converges to $c=1$ with the increase of $L_x$, although it deviates notably from $c=1$ on short cylinders due to the finite-size effect. Therefore, our results on the $L_y=4$ cylinders is consistent with the LE liquid.

%==Summary and discussion==
{\bf Summary and discussion: }%
Our results show that doping the CSL on the triangular lattice can naturally give rise to topological superconductivity with $d\pm id$-wave pairing symmetry. The SC correlations are dominant in the lightly-hole doped region $0<\delta\leq 8.33\%$. Although the CDW correlation also decays with a power-law, it appears secondary since $K_c > K_{sc}$. The extracted $K_{sc}<1$ provides compelling evidences for robust topological superconductivity with divergent SC susceptibility. To the best of our knowledge, this is the first numerical realization of dominant topological $d\pm id$ superconductivity on the triangular lattice in doping a CSL. Our results provide direct supports on theory that doping a CSL or fractional quantum Hall state will yield exotic superconductivity.\cite{Laughlin1988,Laughlin1988a,Wen1989}

In this work, we mainly focus on the lightly doped case, it will be interesting to study the higher doping case and the consequence of doping distinct phases such as the magnetic and tetrahedral phases on the triangular lattice. It is also important to study the effect of second neighbor hopping which is known to be essential to enhance the superconductivity on the square lattice.\cite{Jiang2018,Jiang2018a,Jiang2019} As the CSL can also be realized on other lattices such as Kagome lattice\cite{Bauer2014,Gong2014}, it will be interesting to see whether the topological superconductivity can be realized in doping these systems as well. Answering these questions may help us to better understand the mechanism of high temperature superconductivity.

%==Acknowledgement==
{\bf Acknowledgement: }%
We are grateful to S. Kivelson, H. Yao and T. Devereaux for insightful discussions. This work was supported by the Department of Energy, Office of Science, Basic Energy Sciences, Materials Sciences and Engineering Division, under Contract DE-AC02-76SF00515. Parts of the computing for this project was performed on the Sherlock cluster.

%%reference

\clearpage

\renewcommand{\theequation}{A\arabic{equation}}
\setcounter{equation}{0}
\renewcommand{\thefigure}{S\arabic{figure}}
\setcounter{figure}{0}
\renewcommand{\thetable}{A\arabic{table}}
\setcounter{table}{0}

\begin{widetext}

\section{Supplemental material}

%\subsection{numerical setup?}

\subsection{Superconducting pair-pair correlation}
We calculate six types of superconducting (SC) pair-pair correlations $\Phi_{aa}$, $\Phi_{ab}$, $\Phi_{ac}$, $\Phi_{bb}$, $\Phi_{bc}$ and $\Phi_{cc}$ for all the systems studied. For each $\Phi_{\alpha\beta}(r)$, we calculate $\Phi(\epsilon)$ with the truncation error $\epsilon$ by keeping corresponding number of SU(2) states $m$ and then use second order polynomial function $\Phi(\epsilon)=\Phi_0+a_1 \epsilon + a_2 \epsilon^2$ to extract $\Phi_0$ in the limit $\epsilon\rightarrow 0$, where $a_1$ and $a_2$ are fitting parameters. Examples of the finite truncation error extrapolation for different systems are shown in \Fig{SM-phi}(a-b). For $L_y=4$ cylinders, all types of the SC correlations show clear power-law decaying behaviour at long distance, however, their amplitudes are quantitatively different since the cylindrical geometry explictly breaks the $C_3$ rotational symmetry. For instance, $\Phi_{bb}$, $\Phi_{bc}$ and $\Phi_{cc}$ are significantly stronger than $\Phi_{ab}$ and $\Phi_{ac}$ as shown in \Fig{SM-phi}b, the $\Phi_{aa}$ correlation, on the other hand, is too small to extract a reliable exponent $K_{aa}^{sc}$. The extracted exponents $K_{\alpha\beta}^{sc}$ are fairly close each other, e.g., $K_{ab}^{sc}$=0.78(7), $K_{ac}^{sc}$=0.77(8), $K_{bb}^{sc}$=0.83(8), $K_{bc}^{sc}$=0.83(8) and $K_{cc}^{sc}$=0.83(8) for $L_y=4$ cylinder of length $L_x=72$ at doping level $\delta=5.55\%$. We have obtained similar results for other $L_y=4$ cylinders.

\begin{figure}[b]
    \centering
    \includegraphics[width=0.7\linewidth]{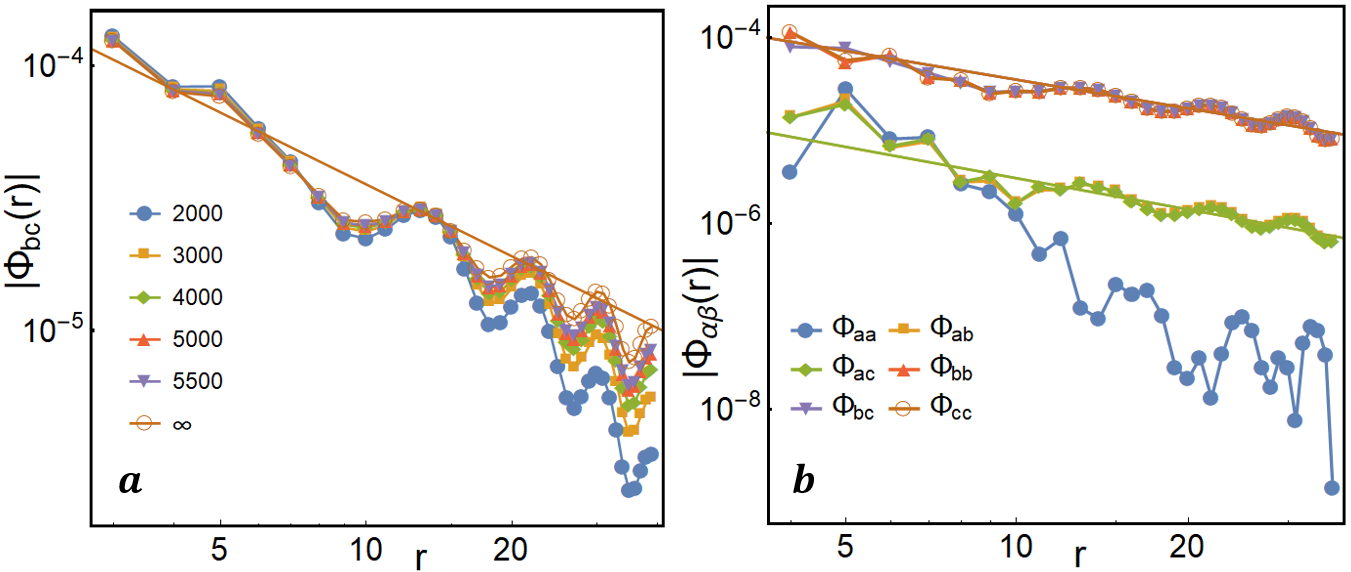}
    \caption{(a) The amplitude of SC correlations $|\Phi_{bc}|$ for $L_y=4$ cylinder of length $L_x=72$ at $\delta=5.55\%$ by keeping different number of SU(2) $m$ states, as well as the extrapolated value in the $m\rightarrow \infty$. (b) The amplitude of all $|\Phi_{\alpha\beta}|$ on cylinder same as (a), extrapolated to the $m\rightarrow \infty$ limit. The solid lines denote power-law fitting to the envelop of $\Phi(r)\sim r^{-K_{sc}}$. All the plots are on the double-logarithmic scale.} \label{SM-phi}
\end{figure}

%For $L_y=5$ cylinders, all SC correlations also decay with a power-law at long distances, however, they have similar amplitude as shown in \Fig{SM-phi}d. The extracted exponents from a cylinder of length $L_x=60$ at doping level $\delta=6.67\%$ (see \Fig{SM-phi}d) are $K_{aa}^{sc}$=1.4(2), $K_{ab}^{sc}$=1.4(1),  $K_{ac}^{sc}$=1.4(1), $K_{bb}^{sc}$=1.5(1), $K_{bc}^{sc}$=1.2(1) and $K_{cc}^{sc}$=1.5(1). We have obtained similar results for other $L_y=4$ and $L_y=5$ cylinders.

We determine the pairing symmetry of the SC correlations by measuring their associated phases $\theta_{\alpha\beta}(r)$, i.e., $\Phi_{\alpha\beta}(r)=\abs{\Phi_{\alpha\beta}(r)}e^{i\theta_{\alpha\beta}(r)}$. To further support the $d\pm id$ pairing symmetry, we have measured $\theta_{\alpha\beta}$ for all the types of the SC correlations. Unlike the Luttinger exponents $K^{sc}$, the phases $\theta_{\alpha\beta}$ are quite stable even with the lowest number of SU(2) states $m=2000$ that we have kept. This is true for both $L_y=4$ and $L_y=5$ cylinders, which also appear true for $L_y=6$ cylinders. Therefore, we can determine the pairing symmetry of the SC correlations on the wider $L_y=6$ cylinders.
%obtain the reliable $\theta_{\alpha\beta}$ for the wider $L_y=6$ cylinders (with next nearest neighbor hopping $t'=1$ used to stabilize the stripes). 
The measured phases $\theta_{\alpha\beta}(r)$ for $L_y=4\sim6$ cylinders are shown in \Fig{SM-phase}(a-c), which provide stronger evidences on the $d\pm id$-wave pairing symmetry as $L_y$ is wider. Interestingly, we also notice that the phases on $L_y=4$ cylinders have more features than the pure $d\pm id$ pairing symmetry, where an additional period-2 oscillation with sign change for $\Phi$ has been observed as shown in \Fig{SM-phase}d. This can be attributed to the period-2 oscillation of spin-spin correlation which was observed only on $L_y=4$ cylinders. To determine the pairing symmetry, we hence introduce a modified phase $\tilde{\theta}(r)=\theta(r)+r\pi$ to subtract such an oscillation. The resulting phases $\tilde{\theta}(r)$, which are shown in \Fig{SM-phi}a and the main text, are consistent with the $d\pm id$-wave pairing symmetry.

\begin{figure}[tb]
    \centering
    \includegraphics[width=\linewidth]{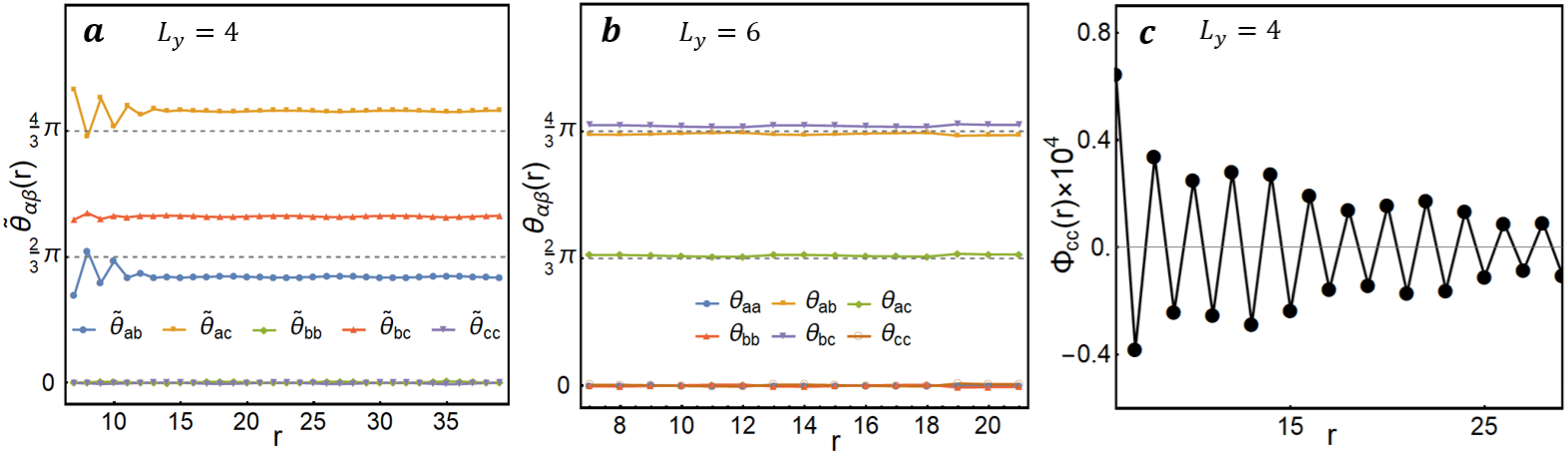}
    \caption{The pattern of phase $\tilde{\theta}_{\alpha\beta}$ associated with $\Phi_{\alpha\beta}$ on (a) $L_y=4$ cylinder of length $L_x=72$ at $\delta=5.55\%$ and (b) $L_y=6$ cylinder of length $L_x=36$ at $\delta=5.55\%$. (c) The SC correlation for $L_y=4$ cylinder with additional period-2 oscillation.}
    \label{SM-phase}
\end{figure}

\subsection{Single-particle correlation}
We have also calculated the single-particle correlation $G(r)=\frac{1}{L_y}\sum_{y=1}^{L_y}\sum_\sigma |\avg{c^\dagger_{x_0,y} c_{x_0+r,y}}|$ for $L_y=4$ cylinders at different doping levels $\delta$
% (see \Fig{SM-sp}).
, where $(x_0,y)$ is the reference site with $x_0\sim L_x/4$ and $r$ is the distance between two sites along the cylinder. At long distances, we find that $G(r)\sim e^{r/\xi_G}$ decays exponentially with the correlation length $\xi_G\sim 1.9$ independent of doping $\delta$. Therefore, there is a finite single-particle gap.

%\begin{figure}[tb]
%    \centering
%    \includegraphics[width=\linewidth]{FigS3new_singleparticle.png}
%    \caption{Single-particle correlation $G(r)$ on $L_y=4$ cylinder of length $L_x=72$ at doping level $\delta=4.17\%$ and $\delta=5.55\%$ on the semi-logarithmic scale. The solid lines denote exponential fitting $G(r)\sim e^{-r/\xi_s}$, with correlation length $\xi_G\sim 1.9$ and $\sim 1.8$.}
%    \label{SM-sp}
%\end{figure}

\begin{figure}[tb]
    \centering
    \includegraphics[width=0.7\linewidth]{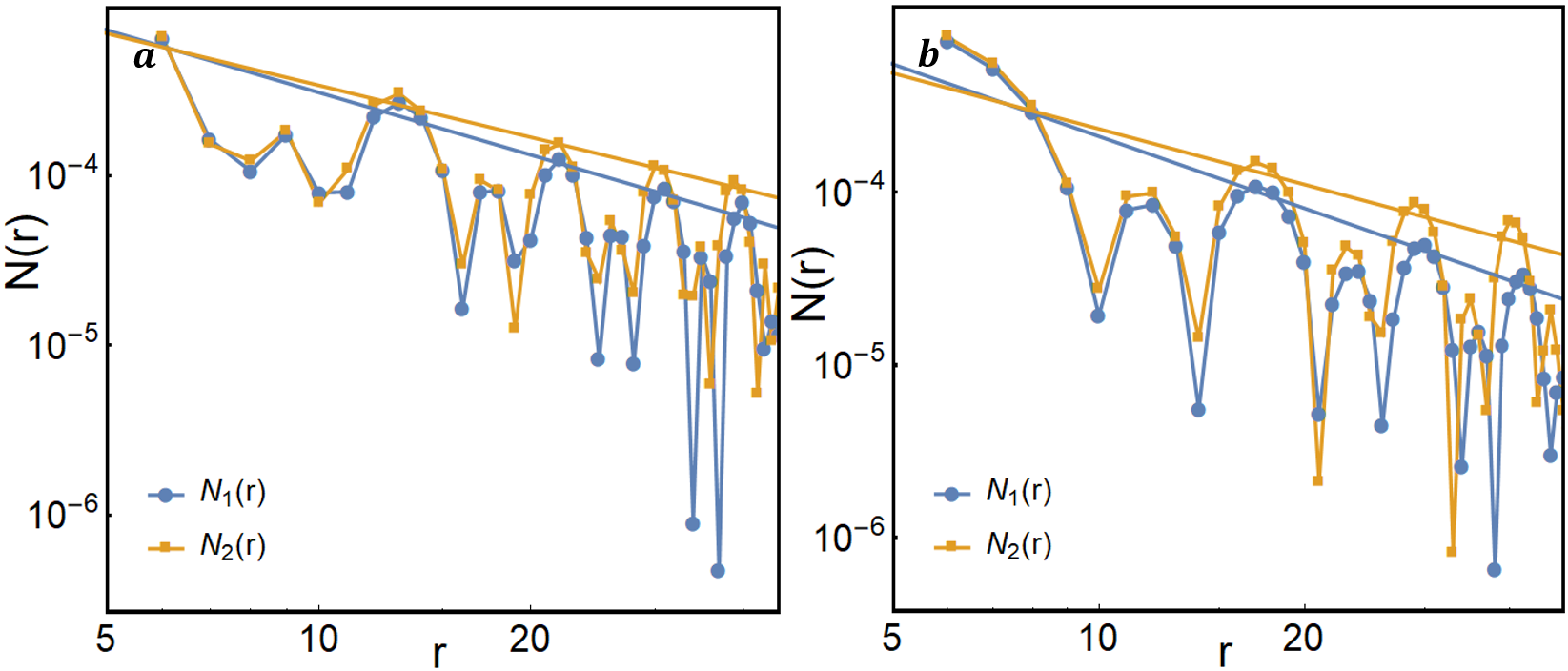}
    \caption{The charge density-density correlations $N_1(r)$ and $N_2(r)$ on $L_y=4$ cylinders of length (a) $L_x=90$ at doping $\delta=5.55\%$ and (b) $L_x=96$ at doping $\delta=4.17\%$. The solid lines denote power-law fitting to the envelop of the corresponding correlations.}
    \label{SM-ncor}
\end{figure}

\subsection{Charge density-density correlation}
In addition to the Friedel oscillation of the charge density profile, the Luttinger exponent $K_c$ can also be obtained from the charge density-density correlations defined by%
\bea
N_1(r)&=&\frac{1}{L_y}\sum_{y=1}^{L_y} \avg{(n_{x_0,y}-\avg{n_{x_0,y}}) (n_{x_0+r,y}-\avg{n_{x_0+r,y}})},\nn \\
N_2(r)&=&\frac{1}{L_y}\sum_{y=1}^{L_y} \avg{(n_{x_0,y}-\bar{n}) (n_{x_0+r,y}-\bar{n})},
\label{SM-eq-N}
\eea
where $\bar{n}$ is the average electron density. Here $(x_0,y)$ is the reference site with $x_0\sim L_x/4$ and $r$ is the distance between two sites along the cylinder. For the quasi-long-range charge order, both $N_1$ and $N_2$ should decay with a power-law in large $r$ limit with the same exponent $K_c$ as determined from the Friedel oscillation. As an example, both $N_1(r)$ and $N_2(r)$ for $L_y=4$ cylinders at two different doping levels are shown in \Fig{SM-ncor}. For $\delta=5.55\%$, $N_1(r)$ and $N_2(r)$ gives $K_c=1.22(9)$ and $K_c=1.02(8)$ respectively, which are quite close to the value obtained from Friedel oscillation $K_c=1.18(5)$. For $\delta=4.17\%$, $N_1(r)$ and $N_2(r)$ gives $K_c=1.4(2)$ and $K_c=1.1(2)$, respectively, which are also consistent with the value from the Friedel oscillation $K_c=1.27(2)$. All these are consistent with the quasi-long-range charge-density-wave (CDW) order.

\begin{figure}[tb]
    \centering
    \includegraphics[width=0.6\linewidth]{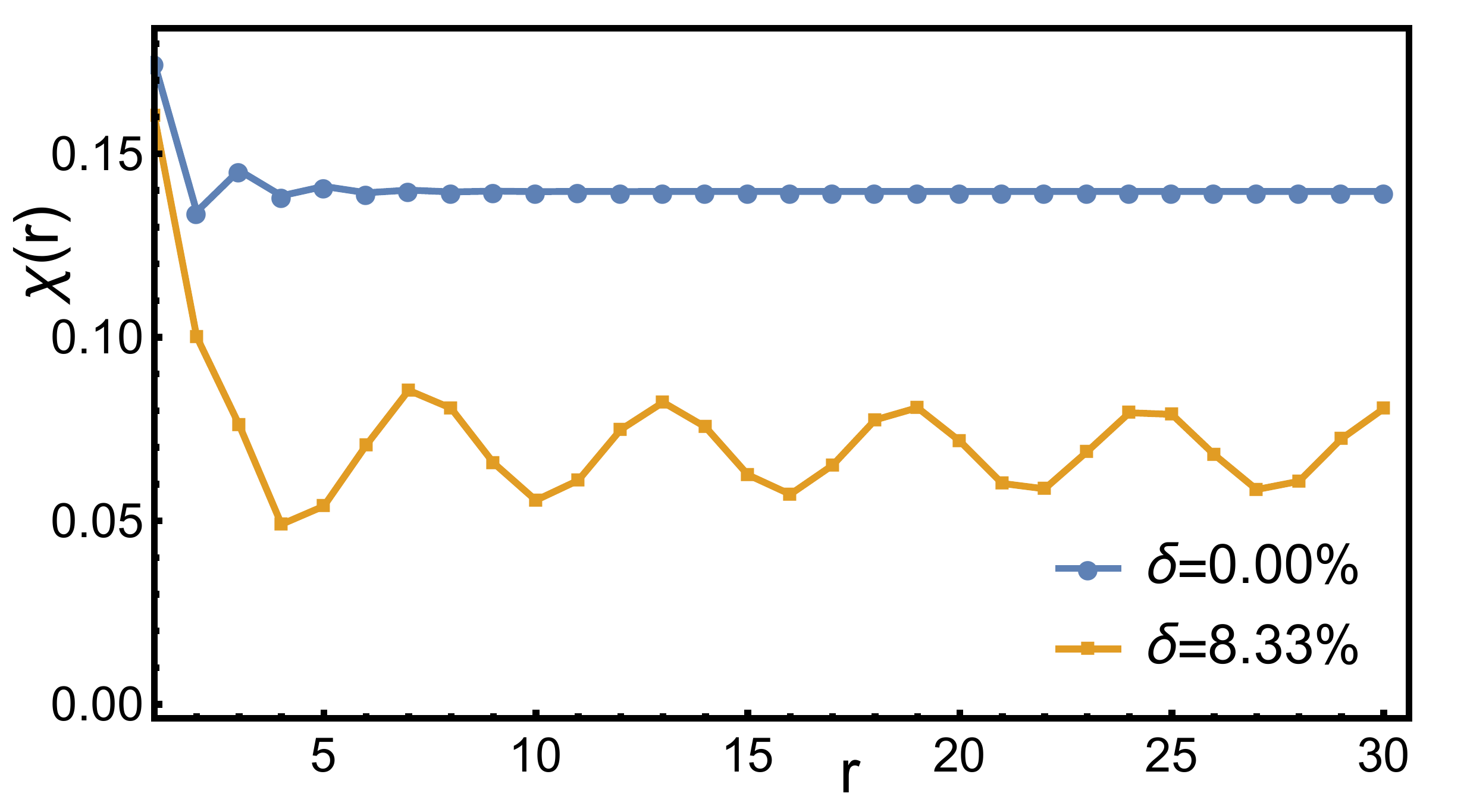}
    \caption{Chiral order parameter $\chi(r)$ measure from the boundary of the $L_y=4$ cylinder of length $L_x=60$ at half filling and doping $\delta=5.55\%$.}
    \label{SM-chiral}
\end{figure}

\subsection{Chiral order}
We measure the chiral order parameter $\chi(r)=\avg{\v{S}_i\times \v{S}_j)\cdot \v{S}_k}$ defined on the upper triangle from the open boundary to the bulk of the cylinder for $J_\chi=0.4$ shown in \Fig{SM-chiral}. For the undoped CSL, the chiral order rapidly converges to finite value consistent with previous study. At finite doping, $\chi(r)$ converges to smaller value combined with additional spatial modulation induced by the charge density wave.

\end{widetext}

\end{document}